\documentstyle[aps,preprint]{revtex}
\input epsf
\tightenlines
\begin{document}

\title{
Superconductor-to-Spin-Density-Wave Transition in Quasi-One-Dimensional 
Metals with Ising Anisotropy
}
\author{A.V. Rozhkov$^{\rm a}$ and A.J. Millis$^{\rm b}$}

\address{
$^{\rm a}$Center~for~Materials~Theory,
Department~of~Physics~and~Astronomy, Rutgers~University,
136~Frelinghuysen~Road,
Piscataway, NJ~08854,~USA
\newline
$^{\rm b}$ Columbia~University, Department~of~Physics, 538~W~120th~Street,
New~York, NY~10027
}

\maketitle

\begin{abstract}
We study a mechanism for superconductivity in quasi-one-dimensional
materials with Ising anisotropy. In an isolated chain Ising anisotropy
opens a spin gap; if inter-chain coupling is sufficiently weak, single
particle hopping is suppressed and the physics of coupled chains is
controlled by a competition between pair hopping and exchange interaction.
Spin density wave and triplet superconductivity phases are found separated
by a first order phase transition. For  particular parameter values a
second order transition described by SO(4) symmetry is found.
\end{abstract}
\hfill
\draft

\section{Introduction}
This paper study a mechanism of superconductivity for
quasi-one-dimensional (Q1D) metals. Well known examples of Q1D materials
exhibiting superconductivity are organic compounds
(TMTSF)$_2$X and (TMTTF)$_2$X. At this moment there is no consensus on
the origin of the superconductivity this metals. Two possibilities are
discussed in the literature. One is BCS theory with the attraction mediated
by either phonons or magnons \cite{book,bour}. 
The second approach developed by
early workers in the field tries to accommodate the
`g-ology' picture of a one-dimensional conductor 
to the materials of interest \cite{brazovski}. The crystal structure of the
materials allows one to treat them as an array of the weakly coupled
one-dimensional conductors formed by chains of flat molecules. It is
assumed further that at hight temperature every conductor can be 
approximately described as a Luttinger liquid (LL) characterized by
dimensionless parameters ${\cal K}_{\rm c,s}$. Depending on those
parameters, at least one of four susceptibilities, spin-density, 
charge-density,
singlet Cooper pair or triplet Cooper pair, diverges at low
temperature. In a three-dimensional crystal a phase with a broken symmetry 
corresponding to the most divergent susceptibility will be formed 
at low temperature.
Thus, the efforts were invested into the study of the LL
parameters as a function of the temperature and the pressure with the hope
of finding conditions under which the Cooper pair susceptibility
becomes divergent.
Due to Coulomb repulsion, however, it was proven to be difficult to
move the system into the region of divergent Cooper pair 
susceptibility \cite{barisic}.

The inability of the `g-ology' picture to offer an explanation for the
superconducting phase, by contrast to its success in describing the
spin-density wave (SDW) phase \cite{nmrI,nmrII} prompted many researchers 
to think in terms of a hybrid approach \cite{book}. 
It is known that in the majority of Q1D materials
SDW and Pierls-SDW phases are stable at ambient pressure. In order to
stabilize superconductivity external pressure has to be applied.
The pressure acts to increase the transverse coupling between chains driving
the system closer to an anisotropic 3D Fermi liquid. It was suggested that
this 3D Fermi liquid is a crucial requirement for the superconductivity 
while the SDW origin is essentially one-dimensional. 
With this agreed upon one
can use BCS theory and its strong-coupling modification to derive different
experimentally relevant quantities.

In this paper we will not follow this path. Instead, it will be
demonstrated that both SDW and superconductivity can be derived within the
framework of weakly coupled one-dimensional conductors. 
Our proposal is very similar in spirit, although different in detail
to that of Carlson {\it et. al.} \cite{erika}
who studied consequences of stripe formation in models of
the cuprate superconductors. 

In its current state the mechanism discussed in this paper cannot
be applied directly to the organic superconductors because it requires
substantial Ising anisotropy whereas in the organic materials deviations
from the rotational symmetry are very small. Yet, it may be a useful
step toward constructing viable mechanism for this materials. We discuss
possible modifications at the end of the paper. It is
also can be considered as a guide in the search for new superconducting
materials. In addition to that it is interesting from purely theoretical
point of view.

There are two parts to the proposed mechanism. 
At high energies inter-chain couplings can be
neglected and the physics is purely
one-dimensional. Unlike the `g-ology' picture, however, it is accepted
that chains
are not simple Luttinger liquids in high energy regime. Instead, we assume
that an interaction which is anisotropic in spin space opens a gap
$\Delta_{\rm s}$ in the spin sector.

Opening of the
gap dramatically changes the low energy behavior of the system. For
reasonable values of $\cal K_{\rm c}$ ($1/2<{\cal K}_{\rm c}<1$)
there are now two relevant operators \cite{gia_schu}. One
is longitudinal SDW (SDW$_z$), the other is triplet superconductivity (TSC)
dual to SDW. Provided
that inter-chain transversal coupling is smaller then $\Delta_{\rm s}$ 
an effective
low energy hamiltonian can be easily derived. It describes the competition
of this two order parameters. The intra-chain electron-electron repulsion
makes $\cal K_{\rm c}$ smaller then unity. In this situation SDW is more
relevant then TSC. Yet, under pressure, the influence of the 
next-to-nearest-neighbor coupling increases. This increases the effective 
Josephson coupling between the chains and decreases the effective exchange
coupling. Therefore, it is possible to observe superconductivity under
pressure. 

An essential feature of the approach is its many-body character. Its main
ingredient is severe spin-charge separation. Unlike the usual BCS
mechanism, this route to
superconductivity does not require attractive interaction.

Another interesting property of the mechanism is that for
some parameter values the symmetry group of the low energy effective
hamiltonian is enhanced.
Generically, the symmetry is U(1)$\otimes$U(1), one copy of U(1) for 
the gauge transformation and another one for translation along the chains. 
If parameters of the hamiltonian are fine-tuned the symmetry group is 
enlarged to SO(4).
An extra symmetry is particle-hole rotation connecting SDW and TSC ground
states.

The rest of the paper is organized as follows. In Section II we introduce
high energy single-chain hamiltonian and discuss its spectrum and
correlation functions. We introduce our low-energy hamiltonian in Section
III. The phase diagram is calculated in Section IV. The effect of long
range Coulomb interaction is considered in Section V. The SO(4) symmetry is
derived in Section VI. Finally, Section VII contains discussion.

\section{High energy description of the system}

It is usually agreed upon that high energy physics of Q1D materials is
purely one-dimensional. The hamiltonian of a single chain can be written as
follows:
\begin{eqnarray}
H&=&\int_{-L}^{L} dx  {\cal H}, \label{H1chain}\\
{\cal H}&=&{\cal H}_{\rm c} + {\cal H}_{\rm s} \\
{\cal H}_{\rm c}&=&\frac{v_{\rm c}}{2}\left({\cal K}_{\rm c}^{-1}\left( 
\nabla\Phi_{\rm c} \right)^2 + 
{\cal K}_{\rm c}^{\vphantom{-1}} 
\left( \nabla\Theta_{\rm c}\right)^2\right)
\label{Hc}\\
{\cal H}_{\rm s}&=&\frac{v_{\rm s}}{2} \left({\cal K}_{\rm s}^{-1}
\left(\nabla\Phi_{\rm s} \right)^2 + 
{\cal K}_{\rm s}^{\vphantom{-1}} 
\left( \nabla\Theta_{\rm s}\right)^2\right) +
\frac{v_{\rm s}g_{\rm bs}}{\pi a^2} \eta_{\rm L\uparrow} 
\eta_{\rm R\uparrow}\eta_{\rm R\downarrow} \eta_{\rm L\downarrow}
\cos\sqrt{8\pi}\Phi_{\rm s}, \label{Hs}
\end{eqnarray}
where we used the Abelian bosonization prescription \cite{boson}:
\begin{eqnarray}
\psi^\dagger_{p\sigma} (x) = 
(2\pi a)^{-1/2} \eta_{p\sigma}{\rm e}^{{\rm i}\sqrt{2\pi} 
\varphi_{p\sigma}(x)}=
(2\pi a)^{-1/2} \eta_{p\sigma}{\rm e}^{{\rm i}\sqrt{\pi/2} \left[
\Theta_{\rm c}(x) + p\Phi_{\rm c}(x) + \sigma\left( \Theta_{\rm s}(x) + 
p\Phi_{\rm s}(x) \right)\right]},\label{bos}
\end{eqnarray}
to express the electron hamiltonian in terms of bosonic fields.
In the above formulae
$\eta_{p\sigma}$ are Klein factors, $\Theta_{\rm s,c}$ are spin and
charge boson fields, $\Phi_{\rm s,c}$ are dual spin and charge fields.
The index $p=\pm$ characterizes the electron direction of motion,
`+' corresponds to `L(eft)'-moving electrons and `-' corresponds to
`R(ight)'-moving electrons. 
At high energy the spin and charge bosons are decoupled.
The parameters of the hamiltonian (\ref{H1chain}) are phenomenological
quantities, functions of the microscopic electron interactions and
dispersion.

Equation (\ref{Hc}) implies that the spectrum of charge excitations is
gapless. The
presence or the absence of the gap in the spin sector is a subtler matter,
and depends on the relation between $g_{\rm bs}$ and $\cal
K_{\rm s}$. A renormalization group (RG) analysis of (\ref{Hs}) leads to
\cite{boson,SG}:
\begin{eqnarray}
&&\frac{d{\cal K}_{\rm s}}{d\ell} = -2g_{\rm bs}^2,
\label{RG1} \\
&&\frac{dg_{\rm bs}}{d\ell} = -2g_{\rm bs}
\left({\cal K}_{\rm s} -1\right)\label{RG2}.
\end{eqnarray}
The infrared properties in the spin sector are defined by the value of the
RG flow invariant:
\begin{equation}
I =g_{\rm bs}^2-\left({\cal K}_{\rm s} -1\right)^2.
\end{equation}
For positive $I$ the cosine term is relevant and the spin bosons spectrum
has a gap.
For $I\le 0$ the cosine term scales down to zero and the spectrum is
gapless. Physically, the value of $I$ corresponds to the spin rotational 
symmetry of the interaction. If the interaction is easy axis (plane) 
then $I>0$ ($I<0$). 

Let us now assume that the interaction has an easy axis. In
this case there is a gap
\begin{equation}
\Delta_{\rm s} \propto \Lambda \exp \left( -\frac{\pi}{2} I^{-1/2}
\right) \label{gap}
\end{equation} 
in the spin sector. Here $\Lambda$ is the ultraviolet cut-off energy.

Opening of the spin gap does not imply long-range ordering. All expectation
values disallowed by symmetries of the hamiltonian (\ref{H1chain}) remain
zero. For example, the operator density corresponding to SDW order takes
the form:
\begin{eqnarray}
{\cal S}_z(2k_{\rm F}) = \frac{1}{2}\left(\psi^\dagger_{\rm L\uparrow}
\psi^{\vphantom{\dagger}}_{\rm R\uparrow} -
\psi^\dagger_{\rm L\downarrow} \psi^{\vphantom{\dagger}}_{\rm R\downarrow}
\right) \propto {\rm e}^{{\rm i}\sqrt{2\pi} 
\Phi_{\rm c}} \sin \sqrt{2\pi} \Phi_{\rm s}. \label{Sz}
\end{eqnarray}
The non-vanishing spin gap fixes the field $\sqrt{2\pi} \Phi_{\rm s}$ 
near $\pm \pi/2$ so the expectation value 
$\langle \sin \sqrt{2\pi} \Phi_{\rm s} \rangle$ is non-zero. The SDW order
parameter, however, is zero because charge boson field $\Phi_{\rm c}$
fluctuates: $\langle {\cal S}_z \rangle \propto \langle \exp ({\rm i}
\sqrt{2\pi} \Phi_{\rm c} ) \rangle = 0$. 
The correlation function $\langle {\cal S}_z
{\cal S}_z \rangle \propto \langle \exp ({\rm i}\sqrt{2\pi} 
\Phi_{\rm c}(x,\tau)) \exp(-{\rm i}\sqrt{2\pi} \Phi_{\rm c}(x', \tau')) 
\rangle$ decays algebraically with the exponent ${\cal K}_{\rm c}$
for large separations.

The state with the spin gap was proposed and described first in 
Ref. \cite{gia_schu}.

\section{Interchain Coupling for $\omega < \Delta_{\rm s}$}

In this section we determine the effect of inter-chain coupling on the
assumption $\omega < \Delta_{\rm s}$.
The hamiltonian density which incorporates such interactions is:
\begin{eqnarray}
{\cal H} = \sum_i {\cal H}_i + \sum_{ij} {\cal H}_{\perp ij}
\label{Htot}\\
{\cal H}_{\perp ij} = t_{\perp ij} \sum_{\sigma} 
\left(\psi^\dagger_{{\rm L}\sigma i}
\psi^{\vphantom{\dagger}}_{{\rm L}\sigma j} +
\psi^\dagger_{{\rm R}\sigma i}
\psi^{\vphantom{\dagger}}_{{\rm R}\sigma j} \right)
+ {\rm h.c.}, \label{Hperp}
\end{eqnarray}
where $i$ and $j$ labels chains and the single chain hamiltonian density 
(\ref{H1chain}) is
${\cal H}_i$. The single particle inter-chain hopping is represented by
the hamiltonian density ${\cal H}_{\perp ij}$.
It will be demonstrated below that the other physically important 
interaction the inter-chain density-density repulsion is irrelevant.

An addition of a single electron or hole to a chain with a spin gap creates
a state with a spin boson soliton. The energy of such soliton is at least
$\Delta_{\rm s}$. Yet, it is possible to add a pair of particles or a
particle and a hole in such a way that the soliton is not created and the
spin bosons are left undisturbed. This implies that while the single
particle hopping in our situation is greatly diminished, the
particle-particle (Cooper pair) hopping and the particle-hole (exchange)
hopping survive the presence of the spin gap \cite{erika,tsvelik_carr}. 
Technically,
when transverse hopping is smaller then $\Delta_{\rm s}$ the fast modes 
(those, whose energy exceeds the spin gap) can
be easily `integrated out'.
The slow degrees of freedom are charge bosons
with the energy less then the spin gap. The effective dynamics of these
modes are given by the following hamiltonian density:
\begin{eqnarray}
{\cal H}_{\rm eff}&=&\frac{v_{\rm c}}{2} \sum_j 
\left({\cal K}_{\rm c}^{-1}\left(\nabla\Phi_{{\rm c}j} \right)^2 +
{\cal K}_{\rm c}^{\vphantom{-1}}
\left( \nabla\Theta_{{\rm c}j}\right)^2\right)
\label{Heff0}\\
&&+2a_{\rm c}^{-1} \sum_{ij} 
\left( J^{\rm sdw}_{ij}
\left(\eta_{{\rm L} \uparrow}^i \eta_{{\rm R} \uparrow}^i 
\eta_{{\rm R}\uparrow}^j \eta_{{\rm L} \uparrow}^j 
+ \eta_{{\rm L} \downarrow}^i \eta_{{\rm R} \downarrow}^i 
\eta_{{\rm R}\downarrow}^j \eta_{{\rm L} \downarrow}^j\right) 
\cos\sqrt{2\pi} \left( \Phi_{{\rm c}i} - \Phi_{{\rm c}j} \right) 
\right.
\nonumber\\
&&\qquad-\left. J^{\rm sc}_{ij}
\left(\eta_{{\rm L} \uparrow}^i \eta_{{\rm R} \downarrow}^i 
\eta_{{\rm R}\downarrow}^j \eta_{{\rm L} \uparrow}^j 
+ \eta_{{\rm L} \uparrow}^i \eta_{{\rm R} \downarrow}^i 
\eta_{{\rm R}\downarrow}^j \eta_{{\rm L} \uparrow}^j\right) 
\cos \sqrt{2\pi} 
\left( \Theta_{{\rm c}i} - \Theta_{{\rm c}j} \right)\right).\nonumber
\end{eqnarray}
Here $J^{\rm sdw}$ is the effective exchange coupling and $J^{\rm sc}$ is
the effective Josephson coupling. The hamiltonian itself is derived in the
Appendix. The ultraviolet cut-off for this effective theory is equal to
$a_{\rm c}^{-1}=\Delta_{\rm s}/v_{\rm c}$.

It is necessary to explain the physical meaning of different terms of
(\ref{Heff0}). The origin of the first term is obvious. 
This term is responsible for the
intra-chain dynamics of the charge bosons. The second term acts to order
$\Phi_{\rm c}$. As one can infer from (\ref{Sz}) the field $\Phi_{\rm c}$ 
should be viewed as the phase of the SDW. 
When $\exp({\rm i}\sqrt{2\pi}\Phi_{\rm c})$ acquires finite expectation
value the ground state becomes SDW$_z$.
One has to remember that $\langle \sin \sqrt{2\pi} \Phi_{\rm s} \rangle$
is non-zero. Unlike ${\cal S}_z$, the
superconducting order parameter in SDW phase is zero.
It is proportional to $\exp\left({\rm i}\sqrt{2\pi} \Theta_{\rm c} \right)$. 
Since $\Theta_{\rm c}$ is dual to $\Phi_{\rm c}$ the ordering of 
the latter implies strong fluctuations of $\Theta_{\rm c}$ and, therefore, 
$\langle\exp\left({\rm i}\sqrt{2\pi} \Theta_{\rm c} \right) \rangle = 0$.

The last term of (\ref{Heff0})
describes the Josephson coupling between pairs of the
chains. If the field $\Theta_{\rm c}$ is ordered the ground state becomes the
longitudinal triplet superconductor. The operator density
\begin{equation}
\Delta_{\pi 0} = \psi^\dagger_{\rm L\uparrow}
\psi^{\dagger}_{\rm R\downarrow} -
\psi^\dagger_{\rm R\uparrow} \psi^{\dagger}_{\rm L\downarrow} \propto
{\rm e}^{{\rm i}\sqrt{2\pi}\Theta_{\rm c}} \sin \sqrt{2\pi} \Phi_{\rm s}
\label{Cooper}
\end{equation}
acquires anomalous non-zero expectation value. We see that the field 
$\Theta_{\rm c}$ plays a role of the superconducting order parameter phase.
The SDW order parameter has zero expectation value in the superconducting
phase.

In a purely one-dimensional system a ground state with a broken
symmetry cannot exist. However, due to inter-chain coupling in
(\ref{Heff0}) the problem at hand becomes three-dimensional where the
broken symmetry phase can be stabilized at low temperature. 
In our situation the longitudinal triplet superconductivity and SDW$_z$ 
are the only types of broken symmetry which are allowed to exist.
Other possibilities, such as
charge-density wave, transversal SDW (SDW$_{x,y}$), longitudinal TSC
or singlet
superconductivity are incompatible with the spin gap and (\ref{Hs}). 
Which of the two allowed ground states, TSC or SDW, has lower energy at $T=0$
depends on the parameters of the effective hamiltonian.

There is a relative minus sign between the second and the third term of
(\ref{Heff0}). This has important consequence. If the ground state is 
a superconductor the order parameter is uniform over the chain array. 
In case of SDW, however, the order parameter has the opposite sign on the
neighboring chains. Therefore, next to nearest neighbor coupling acts to
stabilize the superconductivity and to destabilize SDW. This circumstance
will be used when we will map out the phase diagram of (\ref{Heff0}) in the
next section.

Before we continue with the phase diagram we want to explain why the
inter-chain density-density repulsion can be neglected. We have mentioned
above that the charge-density expectation value is zero due to the spin
gap. Therefore, in the first order of the inter-chain density-density
coupling $V_\perp$ the contribution of this term is exactly zero. In 
the next order the contribution is non-zero. It is proportional to 
$\exp\left(\pm{\rm i}\sqrt{8\pi}\left(
\Phi_{{\rm c} i} - \Phi_{{\rm c}j}\right)\right)$. This operator is
irrelevant and can be neglected.

\section{Phase diagram}

In this section we will determine the phase diagram of the system as a
function of $J$'s and temperature.
Before we start constructing the diagram let us discuss possible
constraints on the values of the model parameters.
There are several parameters our hamiltonian (\ref{Heff0}) is characterized
by: $a_{\rm c}$, $\cal K_{\rm c}$, $J_{ij}$.
We assume that the 
exchange constants $J^{\rm sdw}$ are equal to the Josephson coupling
constants $J^{\rm sc}$. Why this is a reasonable approximation is
discussed in the Appendix. We will put all $J_{ij}$ equal to zero unless
$i$ and $j$ are nearest neighbors or next-to-nearest neighbors.
It is accepted that the next-to-nearest neighbor
coupling constant $J_2$ is smaller then the nearest neighbor constant
$J_1$: $J_1 > J_2$.
When the interaction is repulsive $\cal K_{\rm c}$
is bigger then one half. For the Hubbard model it has been found that $\cal
K_{\rm c}$ is smaller then two. We will confine $\cal K_{\rm c}$ to this
interval:
\begin{equation}
1/2 < {\cal K}_{\rm c} < 1. \label{Krep}
\end{equation}
Outside of this interval the proposed mechanism cannot work:
for ${\cal K}_{\rm c} < 1/2$ the Josephson coupling is irrelevant.

To map out the phase diagram we will use mean field theory
One constructs the mean-field hamiltonian for the coupled chain
problem by re-writing the inter-chain interaction term:
\begin{eqnarray}
&&4\sum_{ij} J_{ij}a_{\rm c}^{-1} \cos \sqrt{2\pi} \left( \Theta_{{\rm c}i}
- \Theta_{{\rm c} j} \right) = {\cal H}_{\rm MF} + \Delta {\cal H},\\
&&{\cal H}_{\rm MF} = -\frac{2}{a_{\rm c}}\sum_{ij} J_{ij} 
\left( \cos \sqrt{2\pi} \Theta_{{\rm c}i}
\left< \cos \sqrt{2\pi} \Theta_{{\rm c}j} \right>+
\cos \sqrt{2\pi} \Theta_{{\rm c}j}
\left< \cos \sqrt{2\pi} \Theta_{{\rm c}i} \right> \right),\\
&&\Delta{\cal H} = -\frac{2}{a_{\rm c}}\sum_{ij}J_{ij} 
\left( 2\cos \sqrt{2\pi}
\left( \Theta_{{\rm c}i}-\Theta_{{\rm c} j} \right) \right.\\
&&\qquad\qquad\left.-\cos \sqrt{2\pi} \Theta_{{\rm c}i}
\left< \cos \sqrt{2\pi} \Theta_{{\rm c}j} \right> - 
\cos \sqrt{2\pi} \Theta_{{\rm c}j}
\left< \cos \sqrt{2\pi} \Theta_{{\rm c}i} \right> \right) \nonumber
\end{eqnarray}
and neglecting $\Delta{\cal H}$.
The mean-field single-chain hamiltonian density is:
\begin{eqnarray}
{\cal H}_{\rm MF} = \sum_i {\cal H}_{{\rm c}i} - 
\tilde{J}_{-} a_{\rm c}^{-1}
\cos \sqrt{2\pi}\Phi_{{\rm c}i} - \tilde{J}_{+} a_{\rm c}^{-1}
\cos \sqrt{2\pi} \Theta_{{\rm c}i},
\label{HMF}\\
\tilde{J}_{-} = J_{-} \langle \cos 
\sqrt{2\pi} \Phi_{\rm c} \rangle,\\
\tilde{J}_{+} = J_{+} \langle \cos 
\sqrt{2\pi} \Theta_{\rm c} \rangle.
\end{eqnarray}
The expression for $\cal H_{\rm c}$ is given by (\ref{Hc}).
Both cosines are relevant interactions. Therefore, they produce gap
$\Delta_{\rm c}$ in the excitation spectrum of the charge boson. But the
nature of the ground state, superconducting or magnetic, depends on the 
values of $\cal K_{\rm c}$ and the effective coupling constants (`+' for
TSC and `-' for SDW):
\begin{equation}
J_\pm = 2J_1 \left( \cos \delta_y^\pm + \cos \delta_z^\pm \right) \pm 2J_2
\left( \cos \left(\delta_y^\pm + \delta_z^\pm \right) +
\cos \left(\delta_y^\pm - \delta_z^\pm\right) + \cos 2\delta_y^\pm + \cos
2\delta_z^\pm\right).\label{Jpm}
\end{equation}
The quantities $\delta_{y,z}$ are differences of the order
parameter phase on neighboring chains. When $J_2$ is zero all $\delta$'s
are zero, too. Otherwise, they have to be determined to maximize the
effective coupling constants $J_\pm$. Simple calculation shows that
$\delta_{y,z}^\pm$ deviate from zero when $|J_2|$ grows bigger then
$J_1/6$:
\begin{equation}
J_{+} = \cases{
4J_1 + 8J_2 & if $J_2> -J_1/2$,\cr
-4J_2-J_1^2/(3J_2) & if $J_2 < -J_1/2$,\cr}\ \ \ \ 
J_{-} = \cases{
4J_1 - 8J_2 & if $J_2< J_1/2$,\cr
4J_2+J_1^2/(3J_2) & if $J_2 > J_1/2$.\cr}
\end{equation}
Of course, $J_{1,2} > 0$ in our system. We will consider the case of
negative $J_2$ for completeness of the presentation.

We did not put the products of the Klein factors in (\ref{HMF}).
This is permissible as
long as the superconductivity and SDW do not co-exist. Then it is possible
to use the following representation of the Klein factors:
\begin{equation}
\eta_{{\rm L}\uparrow}\eta_{{\rm R}\uparrow} = {\rm i},\ 
\eta_{{\rm L}\downarrow}\eta_{{\rm R}\downarrow} = {\rm i},\ 
\eta_{{\rm L}\uparrow}\eta_{{\rm R}\downarrow} = {\rm i},\ 
\eta_{{\rm R}\uparrow}\eta_{{\rm L}\downarrow} = {\rm i}.
\end{equation}
The
averages of cosine of $\Phi_{\rm c}$ and $\Theta_{\rm c}$ have to be found
self-consistently. From the dimensional analysis:
\begin{equation}
\langle \cos \sqrt{2\pi} \Theta_{\rm c} \rangle \propto \left( 
\frac{\Delta_{\rm c}}{\Delta_{\rm s}} \right)^{1/(2{\cal K}_{\rm c})}.
\end{equation}
The required self-consistency condition reads:
\begin{equation}
\tilde{J}_+ \left( \frac{\Delta_{\rm s}}{\Delta_{\rm c}} \right)^{2 - 
1/(2{\cal K}_{\rm c}) } \propto \Delta_{\rm s}.
\end{equation}
These two equations give the following expression for the gap:
\begin{equation}
\Delta_{\rm c} \propto  \Delta_{\rm s} \left( \frac{J_+}{\Delta_{\rm s}}
\right)^{\frac{1}{2 - 1/{\cal K}_{\rm c}}}.\label{SCgap}
\end{equation}
This is the equation for the gap due to the formation of the 
superconducting ground state. The equation for the SDW gap is:
\begin{equation}
\Delta_{\rm c} \propto  \Delta_{\rm s} \left( \frac{J_-}{\Delta_{\rm s}}
\right)^{\frac{1}{2 - {\cal K}_{\rm c}}}.\label{SDWgap}
\end{equation}
The zero-temperature transition between the superconductivity and SDW 
occurs when these two gaps are the same:
\begin{equation}
\Delta_{\rm s} \left( \frac{J_+}{\Delta_{\rm s}}
\right)^{\frac{1}{2 - 1/{\cal K}_{\rm c}}}
\propto  \Delta_{\rm s} \left( \frac{J_-}{\Delta_{\rm s}}
\right)^{\frac{1}{2 - {\cal K}_{\rm c}}}.\label{T=0trans}
\end{equation}
The critical temperature $T_{\rm c}$ for a particular set of parameters 
equals to the biggest gap:
\begin{equation}
T_{\rm c} = \Delta_{\rm s} \max \left\{ \left( \frac{J_+}{\Delta_{\rm s}}
\right)^{\frac{1}{2 - 1/{\cal K}_{\rm c}}},
\left( \frac{J_-}{\Delta_{\rm s}}
\right)^{\frac{1}{2 - {\cal K}_{\rm c}}} \right\}\label{Tc}
\end{equation}
The equation (\ref{T=0trans}) also gives the position of the tricritical
point where the superconductivity, SDW and the phase without the charge
gap co-exist.

Now it is possible to construct the phase diagram. We will
plot it on ($J_2$, $T$) plane. 
It will be explained later how such diagram can be 
related to the ($p$, $T$) digram measured experimentally. 
One has to fix the value of ${\cal K}_{\rm c}$ in the interval (\ref{Krep})
and $r=J_1/\Delta_{\rm s}$ in the interval $0<r\ll 1$. The value of
$J_2$ is constrained to:
\begin{equation}
-J_1 < J_2 < J_1.
\end{equation}
The phase 
boundaries are determined by (\ref{Tc}). The tricritical 
point is given by (\ref{T=0trans}). It is always located at right-hand side 
($ J_2 > 0$) part of the diagram. This is a consequence of (\ref{Krep}).
The point of $T=0$ transition is located right beneath the
tricritical point. The diagram itself is presented on fig.1. The phase with
no broken symmetry is denoted `Spin gap'.

Usually, experimental results are presented on the ($p$, $T$) plane. In
order to establish connection between fig.1 and the experiment we make
three assumptions: (i) the coupling constants $J_{1,2}$ are decreasing
functions of the transverse lattice constant $b_\perp$;
(ii) $J_2$ is more sensitive to change in $b_\perp$ than $J_1$
in the sense that
$J_2 / J_1$ is a decreasing function of $b_\perp$; (iii) both pressure
and temperature affect $J_{1,2}$ through variation of $b_\perp$; all 
other effects of temperature and pressure are unimportant.
The supposition (i) immediately
implies that $J_{1,2}$ are increasing functions of the pressure and
decreasing functions of the temperature. If we
neglect the dependence of $J_1$ on $b_\perp = b_\perp(p,T)$ at all then (ii)
is trivially satisfied. In such a case by applying pressure at zero
temperature we move the
system from SDW at low values of $J_2$ (ambient pressure) to TSC at
higher values of $J_2$ (elevated pressure). Thus, the structure of the
phase diagram at $T=0$ is the same on both ($p$,$T$) and ($J_2$,$T$)
planes.

At non-zero temperature we have to remember that the couplings $J_{1,2}$
are not independent of temperature. Due to the thermal expansion their 
(bare) values decrease as the temperature grows. To compensate this effect
of the thermal expansion extra pressure is required. Therefore, the
tricritical point moves to the right. This effect is commonly referred to
as re-entrance. It is a well-established feature of the experimental phase
diagram of Q1D organic compounds \cite{nmrII}.

Fig.1 suggests that the critical temperature of TSC grows with the 
pressure. This contradicts the experimental data for the organic
superconductors \cite{book} which show that 
$T_{\rm c}$
decreases for $p>p_{\rm TCP}$ where $p_{\rm TCP}$ is the pressure at the
tricritical point. We will address this issue in Section VII.

Let us study the validity of the mean field approach. We will evaluate the
energy correction introduced by the operator $\Delta{\cal H}$. This
correction has to be smaller then the mean field energy 
$E^{(0)} \propto  LN_{\perp}\Delta_{\rm c}^2/v_{\rm c}$ ($L$ is the length
of chains, $N_\perp$ is the number of chains).
The first order correction
is zero: $\left< \Delta{\cal H} \right>_{\rm MF}=0$. Second order
contribution is given by the formula:
\begin{eqnarray}
E^{(2)} = L \int dxd\tau \left< \Delta {\cal H}
(0,0) \Delta {\cal H} (x,\tau) \right>_{\rm MF},\label{E2MF}\\
\Delta {\cal H} (x,\tau) = {\rm e}^{-\tau \int dx' {\cal H}_{\rm MF} (x')}
\Delta {\cal H}(x) {\rm e}^{\tau \int dx' {\cal H}_{\rm MF} (x')}.
\end{eqnarray}
The problem at hand is reduced to the estimation of the bosonic correlation
functions.
When $x/v_{\rm c}$ or $\tau$ exceed $\Delta_{\rm c}^{-1}$ the integrand of
(\ref{E2MF}) vanishes exponentially fast. Thus, the integration can be
restricted to 
$\Delta_{\rm s}^{-1}<\sqrt{\tau^2 + x^2/v_{\rm c}^2}<\Delta_{\rm c}^{-1}$.
At such small distances the effect of the charge gap is negligible. This
simplifies our task even further: the bosons may be considered to be free.
The evaluation of integral (\ref{E2MF}) gives:
\begin{eqnarray}
E^{(2)} \propto \left(L/v_{\rm c} \right)
\left( \frac{\Delta_{\rm s}}{\Delta_{\rm c}} \right)^{2-2/K_{\rm c}}
\sum_{ij} J_{ij}^2 {\rm e}^{2{\rm i} {\bf q} \cdot \left(
{\bf R}_i - {\bf R}_j \right) } \propto LN_\perp\frac{\Delta_{\rm c}^2}
{{\bar z} v_{\rm c}},\\
{\bar z} = \frac{\left(\sum_{ij}J_{ij} {\rm e}^{{\rm i} {\bf q} \cdot 
\left({\bf R}_i - {\bf R}_j \right) } \right)^2 }
{N_\perp\sum_{ij}J_{ij}^2 {\rm e}^{2{\rm i} {\bf q} \cdot \left(
{\bf R}_i - {\bf R}_j \right) } },\\ 
{\bf q}=\cases{0& for TSC\cr b_\perp^{-1} \left( \pi+\delta_y, \pi+\delta_z
\right)& for SDW\cr}.
\end{eqnarray}
Two-dimensional vector ${\bf R}_i$ shows the position of $i$th chain. The
quantity $\bar z$ can be considered as an effective co-ordination number of
a chain. For models with the nearest neighbor interaction only
and ${\bf q}=0$ the value of $\bar z$ coincides with the co-ordination 
number of a chain. In the above expression for $E^{(2)}$ a constant 
independent of $\Delta_{\rm c}$ has been omitted.
We see now that the condition for
the mean field ground state to be a good approximation is
\begin{equation}
{\bar z}\gg 1.\label{z}
\end{equation}
This condition is satisfied if the number of transverse directions is big
or if the transverse interaction is sufficiently long ranged. Inequality
(\ref{z}) is consistent with findings of \cite{tsvelik_carr} where
corrections for the inter-chain mean field theory were also evaluated.

\section{Effect of long range Coulomb interaction}


The phase diagram on fig.1 is a mean-field result. It ignores the
contribution of the Goldstone mode to the free energy. Such contribution is
unimportant at $T=0$. At $T>0$ those modes get excited and increase the
entropy of the system. 
Due to the entropy of these modes the first order phase transition
line separating TSC and SDW phases bents to the right creating
re-entrance region even on $(J_2/J_1, T)$ diagram.

The reason behind the re-entrant behavior is difference between the
Goldstone mode contribution to the entropy of TSC and SDW phases. 
The long-range
Coulomb interaction opens a gap in the spectrum of the Bogoliubov mode
(Goldstone mode in the superconducting phase) but leaves some SDW sliding
modes
without such gap. Therefore, even if at a given value of $J_{1,2}$
the ground state energy of the
superconducting phase is bigger then that of SDW the entropic contribution
of the sliding mode may trigger the transition into SDW phase as the
temperature grows. On $(p, T)$ phase diagram both thermal expansion
discussed in the previous section and the Goldstone modes work together
forming the re-entrance region.

To study the effect of the long-range Coulomb interaction the following
term has to be added to the total hamiltonian:
\begin{equation}
H_{\rm C} = \frac{v_{\rm c} g_{\rm C}}{4\pi} \sum_{ij} \int dx dx' 
\frac{\nabla \Phi_{{\rm c}i} (x)\nabla \Phi_{{\rm c}j} (x')}
{\sqrt{\left(x-x'\right)^2 + \left( {\bf R}_i - {\bf R}_j \right)^2}} 
\approx v_{\rm c} g_{\rm C} \int \frac{dk_\|}{2\pi} 
\frac{d^2 k_\perp}{(2\pi)^2} \frac{k_\|^2}{k_\|^2 + { k}_\perp^2}
\Phi_{\bf k}^{\vphantom{\dagger}} \Phi_{\bf k}^{\dagger}.\label{Coulomb}
\end{equation}
Here $g_{\rm C} \propto e^2/(v_{\rm c}\varepsilon)$ is the interaction 
constant. Simple power counting suggests that (\ref{Coulomb}) is a marginal
operator. Provided that $g_{\rm C} \ll 1$ the effect of the long-range
Coulomb repulsion on the mean-field gap (\ref{SCgap}) and (\ref{SDWgap})
is negligible. However,
it changes the dynamics of the Goldstone modes significantly. In SDW phase
the long-wavelength hamiltonian of the sliding mode is:
\begin{eqnarray}
H^{\rm sdw}&=&\int\frac{dk_\|}{2\pi} 
\frac{d^2 k_\perp}{(2\pi)^2} b_\perp^2 H^{\rm sdw}_{\bf k},\\
H^{\rm sdw}_{\bf k}&=&\frac{1}{2} \left( v_{\rm c} {\cal K}_{\rm c}
\Pi_{\bf k}^{\vphantom{\dagger}} \Pi_{\bf k}^{\dagger} + \left( 
v_{\rm c}^{\vphantom{-1}} {\cal K}_{\rm c}^{-1} k_\|^2 + 
\frac{v_{\rm sdw}^2}{{\cal K}_{\rm c}
v_{\rm c}} {k}_\perp^2 + 2 v_{\rm c}
g_{\rm C} b_\perp^{-2} \frac{k_\|^2}{k_\|^2 + { k}^2_\perp} \right) 
\Phi_{\bf k}^{\vphantom{\dagger}} \Phi_{\bf k}^{\dagger}\right),\\
&&\Pi_{\bf k} = k_\| \Theta_{\bf k},\\
&&v_{\rm sdw}^2 \propto  J_-\Delta_{\rm s} b_\perp^2 \left< 
\cos \sqrt{2\pi} \Phi \right>^2 \propto  \Delta_{\rm c}^2
b_\perp^2.
\end{eqnarray}
Here we used an approximation $\cos \Phi \approx -\left( \Phi^2/2 \right)
\left< \cos \Phi \right>$ to simplify the transversal coupling. The meaning
of the constant $v_{\rm sdw}$ is the velocity of the sliding mode in the
direction normal to the chains. From this expression the dispersion of the
Goldstone bosons is found to be:
\begin{equation}
\omega^2 = v_{\rm c}^2 k_\|^2 + v_{\rm sdw}^2 { k}_\perp^2 + 
2{\cal K}_{\rm c}
v_{\rm c}^2 g_{\rm C} b_\perp^{-2} \frac{k_\|^2}{k_\|^2 + { k}_\perp^2}
\end{equation}
The structure of this formula becomes more clear if we re-parameterize it:
\begin{eqnarray}
\omega^2 = \left(\frac{v_{\rm c}^2  + v_{\rm sdw}^2 }{2}  +
\frac{v_{\rm c}^2  - v_{\rm sdw}^2 }{2}  \cos 2\phi\right) k^2
 + 2{\cal K}_{\rm c} v_{\rm c}^2 g_{\rm C} b_\perp^{-2} \cos^2 \phi,\\
k^2 = k^2_\| + k_\perp^2,\ \tan \phi = k_\perp/k_\|.
\end{eqnarray}
We see that the gap in the sliding mode spectrum closes at $\phi=\pi/2$,
that is, for mode momenta normal to the chain direction.

Similar procedure for the Bogoliubov mode gives:
\begin{eqnarray}
\omega^2 = \left(\frac{v_{\rm c}^2  + v_{\rm sc}^2 }{2}  +
\frac{v_{\rm c}^2  - v_{\rm sc}^2 }{2}  \cos 2\phi\right) \left( k^2
+ 2 {\cal K}_{\rm c} g_{\rm C} b_\perp^{-2} \right),\\
v_{\rm sc}^2 
\propto  J_+\Delta_{\rm s} b_\perp^2 \left< 
\cos \sqrt{2\pi} \Theta \right>^2 \propto  \Delta_{\rm
c}^2 b_\perp^2.
\end{eqnarray}
Unlike SDW sliding mode, the Bogoliubov mode has a gap for any direction of
its momentum. The difference stems from the fact that for TSC case the
long-range Coulomb interaction is expressed not in terms of the field
variable but rather in terms of the conjugate momentum.
The gap is lowest for $\phi=\pi/2$. Its value is $\Delta_{\rm c} g_{\rm
C}^{1/2}$.


\section{SO(4) transformation}

In this section we will show that at some value of the model parameters the
symmetry group of the hamiltonian (\ref{Heff0}) is enhanced. 
One can observe straightforwardly that (\ref{Heff0}) is
invariant under the global shifts
\begin{eqnarray}
\Theta_{{\rm c}i}(x) \rightarrow \Theta_{{\rm c}i}(x) + c,\label{shifts}\\
\Phi_{{\rm c}i}(x) \rightarrow \Phi_{{\rm c}i}(x) + c.\nonumber
\end{eqnarray}
The first of these equations corresponds to the gauge transformation,
the second corresponds to the translation along the chains. Thus, the
symmetry group of the hamiltonian is ${\rm U}(1)\otimes {\rm U}(1)$.
We now show that if the additional conditions:
\begin{equation}
{\cal K}_c = 1,\ J^{\rm sdw}_{ij} = -(-1)^{i-j} J^{\rm sc}_{ij}.
\label{cond}
\end{equation}
are imposed (where $i$ and $j$ are sites in the 2D bipartite lattice formed
by the chains and $(-1)^{i-j}=1$ if $i$ and $j$ are on the same sublattice
and $-1$ if on opposite sublattices) then the hamiltonian symmetry group
is SO(4). To demonstrate this we construct an explicit set of SO(4)
transformations which leave the hamiltonian invariant. As a first step it
is convenient to re-fermionize the hamiltonian using (\ref{bos}) with
$a_{\rm c}$ instead of $a$. We do this first for the inter-chain
interaction term:
\begin{eqnarray}
{\cal H}_{{\rm eff}\perp}&=&8\pi^2a_{\rm c}\sum_{ij}
\left[J^{\rm sdw}_{ij}
\left(\psi_{{\rm L}\uparrow i}^\dagger
\psi_{{\rm R}\uparrow i}^{\vphantom{\dagger}}
\psi_{{\rm R}\uparrow j}^{{\dagger}}
\psi_{{\rm L}\uparrow j}^{\vphantom{\dagger}}
+ \psi_{{\rm R}\downarrow i}^\dagger
\psi_{{\rm L}\downarrow i}^{\vphantom{\dagger}}
\psi_{{\rm R}\downarrow j}^{{\dagger}}
\psi_{{\rm L}\downarrow j}^{\vphantom{\dagger}}
\right)\right.		\label{Heff_hyb}
\\
&&\left.-J^{\rm sc}_{ij}
\left(\psi_{{\rm L}\uparrow i}^\dagger
\psi_{{\rm R}\downarrow i}^{{\dagger}}
\psi_{{\rm R}\downarrow j}^{\vphantom{\dagger}}
\psi_{{\rm L}\uparrow j}^{\vphantom{\dagger}}
+ \psi_{{\rm R}\uparrow i}^{\vphantom{\dagger}}
\psi_{{\rm L}\downarrow i}^{\vphantom{\dagger}}
\psi_{{\rm L}\downarrow j}^{{\dagger}}
\psi_{{\rm R}\uparrow j}^{{\dagger}}
\right)
\right]{\rm e}^{-{\rm i}\sqrt{2\pi} \left(
\Phi_{{\rm s}i} - \Phi_{{\rm s}j} \right)} + {\rm h.c.}\nonumber
\end{eqnarray}
To show the SO(4) structure of this operator we define a four component
spinor
\begin{equation}
\Psi^\dagger = \left(\Psi_{\rm L}^\dagger\ \Psi_{\rm R}^\dagger \right) = 
\left(\psi_{{\rm L}\uparrow}^{\vphantom{\dagger}}
\ \psi_{{\rm L}\downarrow}^\dagger 
\ \psi_{{\rm R}\uparrow}^{\vphantom{\dagger}}
\  \psi_{{\rm R}\downarrow}^\dagger \right) \label{psi}
\end{equation}
and 2$\times$2 matrices:
\begin{eqnarray}
{\hat g}=\Psi_{\rm L}^{\vphantom{\dagger}} \Psi_{\rm
R}^\dagger =
\left(\matrix{ \psi_{{\rm L}\uparrow}^\dagger \psi_{\rm R
\uparrow}^{\vphantom{\dagger}} & \psi_{{\rm L}\uparrow}^\dagger
\psi_{{\rm R}\downarrow}^{{\dagger}}\cr
\psi_{{\rm L}\downarrow}^{\vphantom{\dagger}}
\psi_{{\rm R}\uparrow}^{\vphantom{\dagger}} &
\psi_{{\rm L}\downarrow}^{\vphantom{\dagger}}
\psi_{{\rm R}\downarrow}^{{\dagger}}\cr}\right).
\end{eqnarray}
With this notation the hamiltonian density ${\cal H}_{\rm eff\perp}$ can be
re-written as:
\begin{equation}
{\cal H}_{\rm eff\perp}=4\pi^2 a_{\rm c}\sum_{ij}
\left( \left(J^{\rm sdw}_{ij} + J^{\rm sc}_{ij}\right)
{\rm tr} {\hat g}_i^{\vphantom{\dagger}}
\sigma_z {\hat g}_j^{\dagger}\sigma_z + 
\left(J^{\rm sdw}_{ij} - J^{\rm sc}_{ij}\right)
{\rm tr} {\hat g}_i^{\vphantom{\dagger}}
{\hat g}_j^{\dagger}\right){\rm e}^{-{\rm i}\sqrt{2\pi}
\left(\Phi_{{\rm s}i} - \Phi_{{\rm s}j} \right)} + {\rm h.c.}\label{Hg}
\end{equation}
We now define a set of transformations which (i) leaves the spin boson part
of (\ref{Hg}) invariant and (ii) acts as SO(4) on matrices $\hat g$. 
To do this we
introduce matrices $\sigma_{\mu\nu}$ which act on the four-component spinor
(\ref{psi}). These matrices are determined by:
\begin{eqnarray}
&&\sigma_{\mu\nu} = 
(-{\rm i}/2)\left[\gamma_\mu, \gamma_\nu\right]\label{sigma},
\\
&&\sigma_{ij} = \sum_{k=1}^3 
\varepsilon_{ijk}\left( \matrix{ \sigma_k & 0\cr
                             0 & \sigma_k \cr}\right),\ 
\sigma_{0i} = \left( \matrix{ \sigma_i & 0\cr
                                  0 & -\sigma_i\cr} \right).\label{sigma2}
\end{eqnarray}
with $\gamma$-matrices of Dirac theory \cite{landau}:
\begin{equation}
\gamma_i = \left(\matrix{0 & -{\rm i}\sigma_i \cr
		  {\rm i}\sigma_i & 0 \cr}\right),i=1,2,3,\ 
\gamma_0 = \left(\matrix{  0 & 1\cr
                           1 & 0\cr} \right),\ 
\gamma_5 = \gamma_0\gamma_1\gamma_2\gamma_3 = \left(\matrix{  -1 & 0\cr
                           0 & 1\cr} \right).
\end{equation}
Consider the following transformation group of the spinor $\Psi$:
\begin{equation}
\Psi' = {\exp}\left({\rm i}\sum_{\mu,\nu=0}^3
\omega_{\mu\nu}\sigma_{\mu\nu}\right)\Psi \label{so4}
\end{equation}
with $\omega_{\mu\nu}$ being $c$-numbers.
The form of (\ref{sigma}) and (\ref{sigma2}) shows that
$\Psi_{\rm L}$ and $\Psi_{\rm R}$ transform independently under
(\ref{so4}):
\begin{eqnarray}
&&\Psi_{\rm L}' = U \Psi_{\rm L}\label{U}\\
&&\Psi_{\rm R}' = V \Psi_{\rm R}\label{V}\\
&&{\exp}\left({\rm i}\sum_{\mu,\nu}\omega_{\mu\nu}\sigma_{\mu\nu}\right) 
= \left(\matrix{U & 0\cr
        0 & V\cr}\right), U,V\ - \ {\rm unitary}.
\end{eqnarray}
Because the two L components of $\Psi$ have the same spin boson part (i.e.
same factors of $\exp(\pm{\rm i}\sqrt{\pi/2}\Phi_{\rm s})$ and
$\exp(\pm{\rm i}\sqrt{\pi/2}\Theta_{\rm s})$) as do the two R components and
the transformation (\ref{so4}) does not mix L and R, the spin boson is
left invariant under (\ref{so4}). Therefore, spin boson operators commute
with this transformation, as (i) claims.

It is routinely proven in introductory courses on the quantum
electrodynamics that
the object $R_\mu=\Psi^\dagger\gamma_\mu \Psi$
transforms as a four-dimensional vector under (\ref{so4}).
Therefore, the transformations (\ref{so4}) comprise SO(4) group. 
Unlike the theory of Dirac equation we do not need to deal with $R_\mu$.
To prove (ii)
it is required to know the action of these transformation on $\hat g$.
As a consequence of (\ref{U}) and (\ref{V}) this action is given
by:
\begin{equation}
{\hat g}' = U{\hat g}V^\dagger.
\end{equation}
On a bipartite lattice a global SO(4) rotation can be defined by specifying
matrices $U_{\rm A}$ and $V_{\rm A}$ on sublattice A and assigning $U_{\rm
B}$ and $V_{\rm B}$ according to:
\begin{equation}
U_{\rm B} = \sigma_z U_{\rm A} \sigma_z,\ 
V_{\rm B} = \sigma_z V_{\rm A} \sigma_z.\label{so4g}
\end{equation}
If equation (\ref{cond}) is obeyed then in (\ref{Hg})
the term proportional to $(J_{ij}^{\rm sdw} + J_{ij}^{\rm sc})$ is not
vanishing only 
for $i$ and $j$ on different sublattices and the term proportional to
$(J_{ij}^{\rm sdw} - J_{ij}^{\rm sc})$ only
for $i$ and $j$ on the same sublattice. So the total expression (\ref{Hg})
is invariant under (\ref{so4g}).

The same re-fermionization procedure applied to $\cal H_{\rm c}$ gives:
\begin{eqnarray}
{\cal H}_{\rm c}&=& \sum_i
\frac{\rm i}{2}\left( {\cal K_{\rm c} + {\rm 1}/K_{\rm c}}
\right) v_{\rm c} \sum_\sigma
 \left( \psi^\dagger_{{\rm L} \sigma i} \partial_x
\psi^{\vphantom{\dagger}}_{{\rm L} \sigma i} -
\psi^\dagger_{{\rm R} \sigma i} \partial_x
\psi^{\vphantom{\dagger}}_{{\rm R} \sigma i} \right)\\
&&- 4\pi\left( {\cal K_{\rm c} - {\rm 1}/K_{\rm c}}\right) 
\left( n_{{\rm L} \uparrow
i} + n_{{\rm L} \downarrow i}\right) \left( n_{{\rm R} \uparrow i} +
n_{{\rm R} \downarrow i} \right) -
\left({\cal K_{\rm c} + {\rm 1}/K_{\rm c}}\right) {v_{\rm s}}\left(\left(
\nabla\Phi_{{\rm s}i} \right)^2 +
\left( \nabla\Theta_{{\rm s}i}\right)^2\right) \nonumber\\
&=& \sum_i\frac{(-\rm i)}{2}\left( {\cal K_{\rm c} + {\rm 1}/K_{\rm c}}
\right) v_{\rm c} \left(\Psi^\dagger_i \gamma_5 \partial_x 
\Psi_i^{\vphantom{\dagger}}\right)
+4\pi\left( {\cal K_{\rm c} - {\rm 1}/K_{\rm c}}\right){\rm tr}
{\hat g}_i^{\vphantom{\dagger}}
\sigma_z{\hat g}_i^{\dagger} \sigma_z \nonumber \\
&&-\left({\cal K_{\rm c} + {\rm 1}/K_{\rm c}}\right) {v_{\rm s}}\left(\left(
\nabla\Phi_{{\rm s}i} \right)^2 +
\left( \nabla\Theta_{{\rm s}i}\right)^2\right). \nonumber
\end{eqnarray}
The first and the last terms are invariant under SO(4)
rotations. The second term is not invariant, but vanishes if (\ref{cond})
holds.

To summarize, for generic parameter values the symmetry group of the
hamiltonian (\ref{Heff0}) is U(1)$\otimes$U(1) but if extra conditions
(\ref{cond}) apply then the symmetry of the low energy hamiltonian is 
SO(4).
The symmetry transformations connect smoothly the 
superconducting ground state and the SDW ground state.
A similar situation takes
place in $U<0$ Hubbard model on a square lattice \cite{auerbach} and in
a certain Q1D system near 
charge density wave-singlet superconductivity transition 
\cite{tsvelik_carr}. In the latter paper the symmetry was investigated
using non-Abelian bosonization technique \cite{boson} and found to be
SU(2).
On phenomenological level our model is an analog of
SO(5) theory of high-T$_{\rm c}$ superconductors proposed by S.-C. Zhang
\cite{so5}. 
The long-range Coulomb interaction destroys this symmetry (as seen, for
example, from (\ref{Coulomb})).

\section{Discussion}

In this paper we proposed a many-body theory of the transition from the
spin-density wave to the superconducting state in Q1D materials. The
essential component of the proposed mechanism is the existence of 
the high energy
spin gap on a single chain. This gap is a many-body phenomena unaccessible
from the mean field theory. At the temperature lower then $\Delta_{\rm s}$
all intra-chain interactions and particle hopping are modified by the
presence of the spin gap. Only the exchange interaction and the Josephson
tunneling survive. The competition of these two determines the
low-temperature phase of the system. Our picture allows us to produce a
phase diagram which is qualitatively similar to that of the organic 
superconductors. In particular, we were able to explain the re-entrant 
region of the phase diagram.

In our approach the superconducting phase is deeply connected to SDW. This
connection manifests itself through the existence of the 
quantum symmetry between the superconducting ground state and SDW ground
state. In any real system this symmetry is violated. Depending
on the violation we have either TSC or SDW as a ground state. As we 
demonstrated, the physical long-range Coulomb interaction further 
damages the symmetry by
modifying the Goldstone spectrum of TSC and SDW in different ways. If this
symmetry can be observed experimentally is unclear.

It is also important to note that the superconducting phase is stabilized
in a system with purely repulsive interactions. The total energy is
reduced because the energy of transverse hopping is smaller in the
superconducting phase. This is in
obvious contrast with BCS theory where the optimization of total energy is
achieved by lowering the potential energy of the electron-electron 
interaction. Even more interesting, in BCS superconductivity the pairing
operator is only marginally relevant while in our system it is always
relevant. Thus, $T_{\rm c}^{\rm BCS} \propto \exp(-1/g)$ where $g$ is 
the coupling
constant. In our case $g\propto J_+/\Delta_{\rm s}$ and $T_{\rm c}$ is 
proportional to some power of $g$. In this sense, the proposed mechanism is
`high-temperature' superconductivity.

A crucial prerequisite for experimental implementation of the mechanism is
a strong Ising anisotropy of the spin-spin interaction. Although, in
the organic superconductors this interaction does posses some anisotropy,
the latter seems to be quite weak \cite{magnetic}. The situation becomes
even more aggravated if one realize that the size of the spin gap 
(\ref{gap}) is exponentially small at weak anisotropy. There are two
possible ways out of this problem. First, one can try to improve the
estimate (\ref{gap}). Second, one may consider the system with the small
value of $\Delta_{\rm s}$. We believe, that the latter is a correct route
toward the realistic description of the Q1D organic metals.
An indication that the regime with small $\Delta_{\rm s}$ is
experimentally relevant comes from dependence $T_{\rm c}=T_{\rm c}(p)$. 
Experiments show that the critical temperature is a decreasing function of
the pressure. The situation can be explained
qualitatively if one assume that at high pressure the system becomes 3D
anisotropic Fermi liquid. Since Fermi liquid with repulsive interaction is
stable against transition into the superconducting phase the
superconducting properties deteriorate under pressure. The spin gap also has
to close before the system can become Fermi liquid. Thus, the compound 
inevitably enters the regime where the spin gap is comparable or smaller
then the transversal interactions and the transversal hopping. Capturing
this regime is a significant theoretical challenge.

The major reason why the system with small $\Delta_{\rm s}$ is difficult to
describe is the presence of the unquenched operator of the 
single-electron transversal hopping. When
expressed in terms of the bosonic fields $\Theta_{\rm s,c}$ and $\Phi_{\rm
s,c}$ this operator has distinct non-local structure: it creates kinks in
both $\Theta_{\rm s,c}$ and $\Phi_{\rm s,c}$ fields. This makes it
impossible to apply directly our method to such systems. However, an
important conclusion can be drawn. Q1D metal without well-developed
spin gap on every chain is close to an array of transversally coupled LL.
This puts us in a situation studied before with `g-ology' approach 
\cite{brazovski}. We already mentioned in Introduction about its failure
to describe the superconducting phase. Our study suggests that the
breaking of the spin-rotational invariance can help recover this phase.
Indeed, the scaling dimension of the spin-density and charge-density wave
susceptibilities is
$d_{\rm dw} = {\cal K}_{\rm c} + {\cal K}_{\rm s}$. That of the
Cooper pair susceptibility (longitudinal triplet and singlet) is
$d_{\rm sc} = 1/{\cal K}_{\rm c} + {\cal K}_{\rm s}$. In case of
full rotational invariance (${\cal K}_{\rm s} = 1$) only one of these two
is smaller then 2. However, if ${\cal K}_{\rm s} < 1$ both can be less then
2 simultaneously. In this situation there are several relevant operators in
the system. As we have seen, at the ambient pressure the ground state is
likely to be a density wave since the exchange interaction is more 
relevant then the Josephson coupling operator
($\varepsilon_{\rm dw} < \varepsilon_{\rm sc}$ if ${\cal K}_{\rm c} < 1$).
At the elevated pressure the superconducting state benefits from
next-to-nearest-neighbor processes and may become the ground state.

Despite the appeal of this discussion we have to issue a warning. 
The arguments we made in the previous passage are based on the knowledge of
LL scaling dimensions of different symmetry-breaking operators. In the
presence of the unquenched single-electron hopping those dimensions may
change in some unknown manner. That is why in this paper we chose to study
the case of well-developed spin gap where the single-electron hopping is
irrelevant.

To conclude, we propose a mechanism of SDW and superconductivity in Q1D
materials. This mechanism stabilizes the triplet superconducting state
without any attraction between the electrons. The effective hamiltonian
possesses approximate SO(4) symmetry which connects TSC and SDW ground
states. Depending on how this symmetry is broken the low-temperature phase
is either TSC or SDW.


\section{Appendix}
In this Appendix we will derive the effective low-energy hamiltonian
(\ref{Heff0}). First we recognize the
fact that in the presence of the well defined spin gap $\Delta_{\rm s}$ the
transverse single-electron hopping is ineffective:
when a electron is added to or subtracted from
a chain it creates a soliton in the field $\Phi_{\rm s}$ whose energy
is at least $\Delta_{\rm s}$. Yet, correlated hopping of two electrons can
survive. For example, an addition of a Cooper pair to a chain does not
induce a kink. This can be easily seen from (\ref{Cooper}): the operator
$\Delta_{\pi 0}$ does not contain exponentials of $\Theta_{\rm s}$ which
create kinks. The same is
true for an addition of a particle-hole pair given by operator (\ref{Sz}).
In fact, these two possibilities are the most relevant inter-chain
processes. Their competition dominates the low-energy properties of our
system.

To obtain the low-energy description we have to eliminate all states whose
energy is higher then $\Delta_{\rm s}$.
A particular version of the elimination procedure we will use here is very
similar to the Schrieffer-Wolf transformation known in the theory of Kondo
effect. The idea is to split the Hilbert spaces of a single chain
into two subspaces. The charge subspace ${\cal W}_{{\rm c}j}$ for the 
chain $j$ is spanned by the vectors 
of the form $\left| \psi
_{{\rm c}j} \right> \left| 0_{{\rm s}j} \right>$ where 
$\left| 0_{{\rm s}j} \right>$ is the ground state of the spin boson.
The effective hamiltonian density on the subspace 
${\cal W}_{\rm c} = \sum_{\oplus\ i} {\cal W}_{{\rm c}i}$ can be defined by
its action on $\left| \psi \right> \in {\cal W}_{\rm c}$:
\begin{eqnarray}
H_{\rm eff}\left| \psi \right> = 
{\cal P}_{\rm c}  \left(\sum_i {H}_{i} - {H}_\perp {\cal P}_{\rm c}^{\perp} 
\left( \sum_j {H}_{{\rm s}j} +
{H}_{{\rm c} j} - E \right)^{-1}{\cal P}_{\rm c}^{\perp} {H}_\perp  \right)
{\cal P}_{\rm c} \left| \psi \right> = E \left| \psi \right>,
\label{Heff}
\end{eqnarray}
Here ${\cal P}_{\rm c}$ is the orthogonal projector on ${\cal
W}_{\rm c}$ and ${\cal P}_{\rm c}^{\perp} = 1 - {\cal P}_{\rm c}$.

Unlike usual hamiltonian (\ref{Heff}) depends on the eigenvalue $E$. We will
demonstrate that this dependence is weak and for our purposes (\ref{Heff})
can be treated as an ordinary hamiltonian.
The expression (\ref{Heff}) is quite general. It is possible to simplify it
by noting that the partial matrix element
$\left< 0_{{\rm s}j} \right| H_\perp \left| 0_{{\rm s}j} \right>$ is zero.
This allows to drop the projectors from formula (\ref{Heff}):
\begin{eqnarray}
H_{\rm eff}\left| \psi_{\rm c} \right> = 
\left(\sum_i {H}_{{\rm c}i} -  \left< 
{H}_\perp\left( \sum_j {H}_{{\rm s}j} +
{H}_{{\rm c} j} - E \right)^{-1} 
{H}_\perp  \right>_{\rm s} \right) 
\left| \psi_{\rm c} \right> = E \left| \psi_{\rm c} \right>,
\label{Heff1}
\end{eqnarray}
where $\left< \ldots \right>_{\rm s}$ denotes the partial matrix element of
an operator with respect to the spin boson ground state. We will calculate
this matrix element below.

It is convenient to re-write the inverse operator as follows:
\begin{eqnarray}
&& \left< {H}_\perp\left( \sum_j {H}_{{\rm s}j} +
{H}_{{\rm c} j} - E \right)^{-1} 
{H}_\perp  \right>_{\rm s}  = 
\int_0^\infty du
\left< {H}_\perp {\rm e}^{ -u \left(\sum_j {H}_{{\rm s}j} +
{H}_{{\rm c} j} - E \right)}
{H}_\perp  \right>_{\rm s}  \label{interaction}\\
&=&\int_0^\infty du
\left< {H}_\perp \left( 0 \right)
{H}_\perp\left( u \right)  \right>_{\rm s}
{\rm e}^{ -u \left( \sum_j H_{{\rm c}j} - E \right)} \approx
\int_0^\infty du\left< {H}_\perp \left( 0 \right)
 {H}_\perp\left( u \right)  \right>_{\rm s},
\nonumber
\end{eqnarray}
where we took into account that, to the lowest order in 
$t_{\perp}$, $\left( \sum_j H_{{\rm c} j} - E \right) \left| \psi_{\rm c}
\right> \approx 0$ and as well as that
$\left< H_\perp (0) H_\perp (u) \right>_{\rm s}$ vanishes exponentially 
for $u>\Delta_{\rm s}^{-1}$.
The calculation of the partial matrix element
$
\left< H_\perp \left( 0 \right) H_\perp \left( u \right) \right>_{\rm s} 
$
can be reduced to the calculation of the Green's function for the spin
bosons:
\begin{eqnarray}
&&\left< H_\perp \left( 0 \right) H_\perp \left( u \right) \right>_{\rm s} =
\int dx dx' \sum_{ij}
\left< {\cal H}_{\perp ij}\left( x, 0 \right) 
{\cal H}_{\perp ij}\left( x', u \right) \right>_{\rm s} \label{green}\\
&&=\frac{1}{(2\pi a)^2}
\int dx dx' \sum_{ij} t_{\perp ij}^2
\sum_{p\vphantom{'}\sigma\vphantom{'}}\sum_{p'\sigma'}  
\eta_{p\sigma}^i \eta_{p\sigma}^j \eta_{p'\sigma'}^j \eta_{p'\sigma'}^i 
\nonumber\\
&&\qquad
\times \left<{\rm e}^{{\rm i}\sqrt{2\pi} \varphi_{p\sigma i}(x,0)}
{\rm e}^{-{\rm i}\sqrt{2\pi}\varphi_{p\sigma j}(x,0)}
{\rm e}^{{\rm i}\sqrt{2\pi}\varphi_{p'\sigma' j}(x',u)}
{\rm e}^{-{\rm i}\sqrt{2\pi}\varphi_{p'\sigma' i}(x',u)} \right.\nonumber \\
&& \qquad\quad\left. - {\rm e}^{{\rm i}\sqrt{2\pi} 
\varphi_{p\sigma i}(x,0)}
{\rm e}^{-{\rm i}\sqrt{2\pi}\varphi_{p\sigma j}(x,0)}
{\rm e}^{-{\rm i}\sqrt{2\pi}\varphi_{p'\sigma' j}(x',u)}
{\rm e}^{{\rm i}\sqrt{2\pi}\varphi_{p'\sigma' i}(x',u)} + {\rm h.c.}
\right>_{\rm s}.\nonumber
\end{eqnarray}
The operator $\varphi_{p\sigma}$ is defined by (\ref{bos}). The first term
in the brackets corresponds to a tunneling event in which an electron and a
hole hop from chain $i$ to chain $j$. The second term corresponds to
tunneling of a Cooper pair (two electrons) from $j$ to $i$.
For every pair of chains $i$ and $j$ there are 64 terms in the expression
(\ref{green}). Yet, only a fraction of them has non-zero partial matrix
element $\left<0_{\rm s}\right| \ldots \left|0_{\rm s}\right>$. For this
matrix element to be non-zero the term must be `neutral' with respect to
kink creation operator $\exp\left({\rm i}\sqrt{\pi/2} \Theta_{\rm s} \right)$.
If at $u=0$ a kink (anti-kink) at $i$th chain is created then for the term to
be `neutral' an anti-kink (kink) has to be created at $u>0$. This condition
immediately reduces the number of possible terms. The surviving terms
have the form $\left<T_i^{\vphantom{\dagger}} T_j^\dagger\right>_{\rm s}$ 
where $T$ is one of the following:
\begin{eqnarray}
\psi_{{\rm L} \uparrow}^\dagger(x,0) \psi_{{\rm R} \uparrow}
^{\vphantom{\dagger}}(x',u),\ 
\psi_{{\rm L} \downarrow}^\dagger(x,0) \psi_{{\rm R} \downarrow}
^{\vphantom{\dagger}}(x',u);\\
\psi_{{\rm L} \uparrow}^\dagger(x,0) \psi_{{\rm R} \downarrow}
^\dagger(x',u),\ 
\psi_{{\rm R} \uparrow}^\dagger(x,0) \psi_{{\rm L} \downarrow}
^\dagger(x',u);\\ 
\psi_{{\rm L} \uparrow}^\dagger(x,0) \psi_{{\rm L} \downarrow}
^\dagger(x',u),\ 
\psi_{{\rm R} \uparrow}^\dagger(x,0) \psi_{{\rm R} \downarrow}
^\dagger(x',u);\\
\psi_{{\rm L} \uparrow}^\dagger(x,0) \psi_{{\rm L} \uparrow}
^{\vphantom{\dagger}}(x',u),\
\psi_{{\rm R} \uparrow}^\dagger(x,0) \psi_{{\rm R} \uparrow}
^{\vphantom{\dagger}}(x',u);\\
\psi_{{\rm L} \downarrow}^\dagger(x,0) \psi_{{\rm L} \downarrow}
^{\vphantom{\dagger}}(x',u),\
\psi_{{\rm R} \downarrow}^\dagger(x,0) \psi_{{\rm R} \downarrow}
^{\vphantom{\dagger}}(x',u).
\end{eqnarray}
On this list only two first lines are of interest to
us since they are the most relevant operators. Let us calculate the matrix
element
\begin{eqnarray}
\left<\psi_{{\rm L} \uparrow i}^\dagger(x,0) \psi_{{\rm L} \uparrow j}
^{\vphantom{\dagger}}(x,0)\psi_{{\rm R} \uparrow j}^\dagger(x',u) 
\psi_{{\rm R} \uparrow i}^{\vphantom{\dagger}}(x',u)
\right>_{\rm s} \label{nonlocal}\\
=\left<\psi_{{\rm L} \uparrow i}^\dagger(x,0) \psi_{{\rm R} \uparrow i}
^{\vphantom{\dagger}}(x',u)\right>_{\rm s}
\left<\psi_{{\rm L} \uparrow j}^{\vphantom{\dagger}}(x,0)
\psi_{{\rm R} \uparrow j}^\dagger(x',u) \right>_{\rm s}  \nonumber \\
=-\frac{1}{4\pi^2 a^2}
\eta_{{\rm L} \uparrow}^i \eta_{{\rm R} \uparrow}^i \eta_{{\rm R}
\uparrow}^j \eta_{{\rm L} \uparrow}^j
{\rm e}^{{\rm i}\sqrt{\pi/2}\left(\Theta_{{\rm c}i}(x,0) + 
\Phi_{{\rm c} i}(x,0)\right)}
{\rm e}^{-{\rm i}\sqrt{\pi/2}\left(\Theta_{{\rm c}i}(x',u) - 
\Phi_{{\rm c} i}(x',u)\right)} \nonumber \\
\times {\rm e}^{{\rm i}\sqrt{\pi/2}\left(\Theta_{{\rm c}j}(x,0) - 
\Phi_{{\rm c} j}(x,0)\right)}
{\rm e}^{-{\rm i}\sqrt{\pi/2}\left(\Theta_{{\rm c}j}(x',u) + 
\Phi_{{\rm c} j}(x',u)\right)} 
\left|{\cal G}_{\rm s} (x-x',u)\right|^2, \nonumber \\
{\cal G}_{\rm s} = \left<0_{\rm s} \left| 
{\rm e}^{{\rm i}\sqrt{\pi/2}\left(\Theta_{{\rm s}i}(x,0) + 
\Phi_{{\rm s} i}(x,0)\right)} 
{\rm e}^{-{\rm i}\sqrt{\pi/2}\left(\Theta_{{\rm s}i}(x',u) - 
\Phi_{{\rm s} i}(x',u)\right)} 
\right| 0_{\rm s} \right>.
\end{eqnarray}
The correlation function $\cal G_{\rm s}$ decays exponentially for
$|x-x'|/v_{\rm s}$ and $|u|$ above $\Delta_{\rm s}^{-1}$.
Unfortunately, the operator (\ref{nonlocal}) is quite complicated due to
its non-local structure. An obvious way to simplify it is to approximate
$\cal G_{\rm s}$ by a delta-function. This step is acceptable for
the low-momentum boson whose wave length is
bigger then $v_{\rm c}/\Delta_{\rm s}$. Yet, the presence of high-momentum
bosons render this procedure unsatisfactory. This obstacle can be
overcome by eliminating high-momentum bosons from the effective theory. The
procedure itself is very similar to the elimination of the spin bosons: in
addition to averaging over the spin boson ground state it is required to
average over the ground states of high-momentum charge bosons. After this
procedure (\ref{nonlocal}) is reduced to:
\begin{eqnarray}
\int dx dx' du \left<\left<
\psi_{{\rm L} \uparrow i}^\dagger(x,0) \psi_{{\rm L} \uparrow j}
^{\vphantom{\dagger}}(x,0)\psi_{{\rm R} \uparrow j}^\dagger(x',u) 
\psi_{{\rm R} \uparrow i}^{\vphantom{\dagger}}(x',u)
\right>\right>_{\rm s,c >} 
\label{local}\\
=-\frac{\eta_{{\rm L} \uparrow}^i \eta_{{\rm R} \uparrow}^i \eta_{{\rm R}
\uparrow}^j \eta_{{\rm L} \uparrow}^j}{4\pi^2 a^2}
\int dx {\rm e}^{{\rm i}\sqrt{2\pi}
\left( \Phi_{{\rm c}i} (x,0) - \Phi_{{\rm c}j} (x,0) \right) } 
\int dx' du \left|{\cal G}_{\rm s} (x-x',u)\right|^2 
\left|{\cal G}_{\rm c}^> (x-x',u)\right|^2,  \nonumber \\
{\cal G}_{\rm c}^>(x,u) 
 = \left<0_{\rm c}^> \left| 
{\rm e}^{{\rm i}\sqrt{\pi/2}\left(\Theta_{{\rm c}i}^>(x,0) + 
\Phi_{{\rm c} i}^>(x,0)\right)} 
{\rm e}^{-{\rm i}\sqrt{\pi/2}\left(\Theta_{{\rm c}i}^>(x',u) - 
\Phi_{{\rm c} i}^>(x',u)\right)} 
\right| 0_{\rm c}^> \right>\\
\propto \left(\frac{\Delta_{\rm s}}{\Lambda}\right)^
{{\cal K_{\rm c}}/2} \left( \Lambda \sqrt{u^2 + x^2/v_{\rm c}^2}
\right)^{\left(1/{\cal K}_{\rm c} - {\cal K}_{\rm c} \right)/4}.\nonumber
\end{eqnarray}
In this formula the superscript `$>$' is used to denote the quantities
associated with the high-momentum charge bosons.
Putting the expression (\ref{local})
back into (\ref{interaction}) and performing the
integration over $u$ and $x-x'$ one gets an inter-chain interaction term
of the form:
\begin{eqnarray}
&&J^{\rm sdw}\eta_{{\rm L} \uparrow}^i \eta_{{\rm R} \uparrow}^i 
\eta_{{\rm R}\uparrow}^j \eta_{{\rm L} \uparrow}^j 
\int a_{\rm c}^{-1} dx {\rm e}^{{\rm i}\sqrt{2\pi}
\left( \Phi_{{\rm c}i} (x,0) - \Phi_{{\rm c}j} (x,0) \right) },\\
J^{\rm sdw}&=&t^2_\perp \frac{a_{\rm c}}{4\pi^2 a^2}
\int dx du \left| {\cal G}_{\rm s}(x,u) {\cal G}_{\rm c}^>(x,u) 
\right|^2\label{Jsdw}\\
a_{\rm c}^{-1} = \Delta_{\rm s}/v_{\rm c},\ 
r=\sqrt{u^2 + x^2/v_{\rm c}^2}.\nonumber
\end{eqnarray}
The evaluation of this integral requires knowledge of ${\cal G}_{\rm s}$ at
the intermediate values of $r$: $\Lambda^{-1}<r<\Delta_{\rm s}^{-1}$. 
This knowledge is unavailable.
To circumvent this problem we neglect any anomalous dimension to ${\cal
G}_{\rm s}$ and approximate it by $(\Lambda r)^{-1/2}$. This gives:
\begin{equation}
J^{\rm sdw} \propto \frac{t_\perp^2}{\Delta_{\rm s}}
\left( \frac{\Delta_{\rm s}}{\Lambda}
\right)^{\left({\cal K}_{\rm c} + 1/{\cal K}_{\rm c}\right)/2 -1}.
\end{equation} 
Other terms can be processed in a similar fashion. The resultant
inter-chain interaction is:
\begin{eqnarray}
H_{\rm eff \perp}& =& \sum_{ij}J^{\rm sdw}_{ij}
\left(\eta_{{\rm L} \uparrow}^i \eta_{{\rm R} \uparrow}^i 
\eta_{{\rm R}\uparrow}^j \eta_{{\rm L} \uparrow}^j 
+ \eta_{{\rm L} \downarrow}^i \eta_{{\rm R} \downarrow}^i 
\eta_{{\rm R}\downarrow}^j \eta_{{\rm L} \downarrow}^j\right) 
\int dx a^{-1}_{\rm c}{\rm e}^{{\rm i}\sqrt{2\pi}
\left( \Phi_{{\rm c}i} - \Phi_{{\rm c}j} \right) }\\
&&-J^{\rm sc}\left(\eta_{{\rm L} \uparrow}^i \eta_{{\rm R} \downarrow}^i 
\eta_{{\rm R}\downarrow}^j \eta_{{\rm L} \uparrow}^j 
+ \eta_{{\rm L} \uparrow}^i \eta_{{\rm R} \downarrow}^i 
\eta_{{\rm R}\downarrow}^j \eta_{{\rm L} \uparrow}^j\right) 
\int dx a^{-1}_{\rm c}{\rm e}^{{\rm i}\sqrt{2\pi}
\left( \Theta_{{\rm c}i} - \Theta_{{\rm c}j} \right) } + 
{\rm h.c}.\nonumber
\end{eqnarray}
The value of $J^{\rm sc}$ can be obtained without any calculation if one
takes the duality transformation of the charge hamiltonian:
$\Theta_{\rm c} \leftrightarrow \Phi_{\rm c}$, ${\cal K}_{\rm c} 
\leftrightarrow 1/{\cal K}_{\rm c}$. The quantity dual to $J^{\rm sdw}$ is
$J^{\rm sc}$. A consequence of (\ref{Jsdw}) and the duality is the
following estimate:
\begin{eqnarray}
J^{\rm sc} \propto
\frac{t_\perp^2}{\Delta_{\rm s}} 
\left( \frac{\Delta_{\rm s}}{\Lambda}
\right)^{\left({\cal K}_{\rm c} + 1/{\cal K}_{\rm c}\right)/2 -1} \propto
J^{\rm sdw}.
\end{eqnarray}
Thanks to the duality, we can make a more accurate statement about these
two coupling constants: $J^{\rm sdw} = F({\cal K}_{\rm c})J^{\rm sc}$,
where the function $F({\cal K}_{\rm c})$ is unity when ${\cal K}_{\rm c}$
is unity. Away from this point $J^{\rm sdw}/J^{\rm sc} = {\cal O}(1)$. 

The cut-off of this effective theory of the
charge bosons is now equal to $a_{\rm c}^{-1}$.

The products of Klein factors do not commute with each other. Thus, in
general, they cannot be dropped. Yet, within an inter-chain mean field
approximation the contribution of the Josephson tunneling term in SDW phase
is zero and, conversely, the contribution of the exchange term in the
superconducting phase is also zero. In such a case, it is permissible to
ignore Klein factors.

\begin{figure} [!t]
\centering
\leavevmode
\epsfxsize=8cm
\epsfysize=8cm
\epsfbox[18 144 592 718] {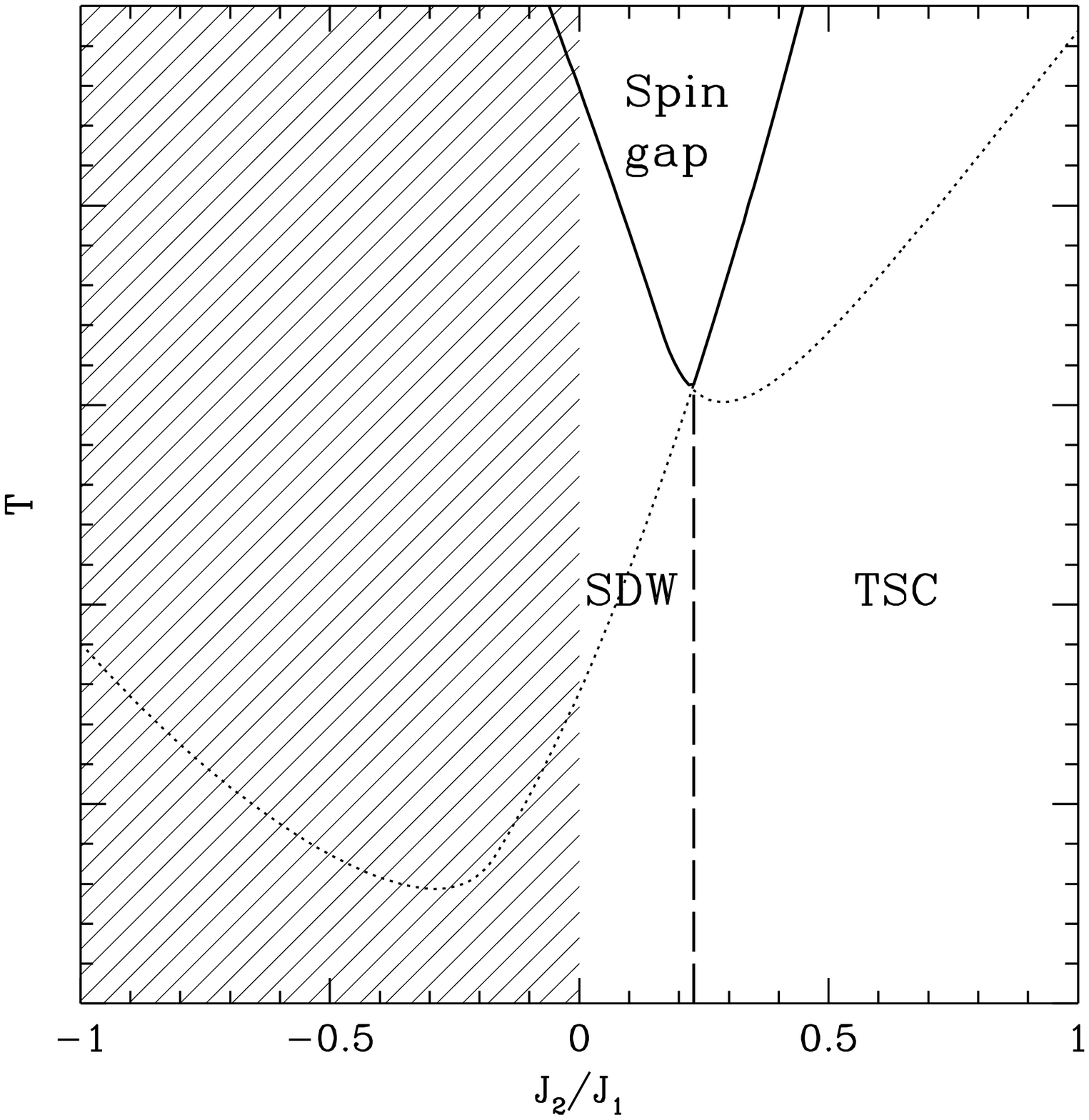}
\caption[]
{\label{fig1} 
Mean-field phase diagram of our model. Solid lines mark the location of the
second-order phase transition into SDW or TSC. Dash line corresponds to the
first order phase transition between SDW and TSC. Label `Spin gap' denotes
the phase in which the charge bosons are disordered. The gap in the charge
sector (equations (\ref{SCgap}) and (\ref{SDWgap})) are shown by dotted
lines. The unphysical region ${J}_2<0$ is shaded.
}
\end{figure}

\end{document}